\documentclass[journal]{IEEEtran}
\usepackage{graphicx}
\PassOptionsToPackage{hyphens}{url}
\usepackage{hyperref}
\usepackage{multirow}
\usepackage{array}
\newcolumntype{C}[1]{>{\centering\arraybackslash}m{#1}}
\usepackage{diagbox}
\usepackage{bbding}
\usepackage{amssymb}
\usepackage{pifont}
\newcommand{\cmark}{\ding{51}}%
\newcommand{\xmark}{\ding{55}}%
\usepackage{xcolor}
\usepackage{array}
\newcolumntype{P}[1]{>{\centering\arraybackslash}m{#1}}

\ifCLASSINFOpdf
\else
\fi
\begin{document}
%
\title{Knowledge-powered Explainable Artificial Intelligence (XAI) for Network Automation Towards 6G}
%
%
%

\author{Yulei~Wu,~\IEEEmembership{Senior Member,~IEEE},
        Guozhi~Lin~and~Jingguo~Ge

\thanks{This work was partially supported by the Engineering and Physical Sciences Research Council (EPSRC) under Grant No. EP/R030863/1. \textit{(Corresponding Author: Yulei Wu)}}
\thanks{Y. Wu is with College of Engineering, Mathematics and Physical Sciences, University of Exeter, Exeter, EX4 4QF, U.K. e-mail: y.l.wu@exeter.ac.uk}
\thanks{G. Lin and J. Ge are with the School of Cyber Security, University of Chinese Academy of Sciences, Beijing, 100049, China and the Institute of Information Engineering, Chinese Academy of Sciences, Beijing, 100093, China. e-mail: linguozhi@iie.ac.cn, gejingguo@iie.ac.cn}
}

%
%

\markboth{}%
{}
%



\maketitle

\begin{abstract}
Communication networks are becoming increasingly complex towards 6G. Manual management is no longer an option for network operators. 
Network automation has been widely discussed in the networking community, and it is a sensible means to manage the complex communication network.
Deep learning models developed to enable network automation for given operation practices have the limitations of 1) lack of explainability and 2) inapplicable across different networks and/or network settings. To tackle the above issues, in this article we propose a new knowledge-powered framework that provides a human-understandable explainable artificial intelligence (XAI) agent for network automation. A case study of path selection is developed to demonstrate the feasibility of the proposed framework. Research on network automation is still in its infancy. Therefore, at the end of this article, we provide a list of challenges and open issues that can guide further research in this important area.

\end{abstract}

\begin{IEEEkeywords}
Network automation, Explainable artificial intelligence (XAI), Human-understandable XAI, 6G, Network management.
\end{IEEEkeywords}

%
\IEEEpeerreviewmaketitle

\section{Introduction}
3GPP completed its Release 16 in June 2020 which is an important milestone for the initial complete 5G system specification. The current plan for the completion of Release 17 is Q1/Q2 2022. According to the technology involvement and the specifications in both Release 16\footnote{\url{https://www.3gpp.org/release-16}} and Release 17\footnote{\url{https://www.3gpp.org/release-17}}, it has been widely agreed that beyond 5G (B5G) and 6G will be facing diversified vertical services. Applications in this vertical include autonomous vehicle, smart factory, remote surgery, intelligent city management, smart grid, and many more. With the fast growth of vertical markets, the complexity of underlying communication networks (e.g., 5G/B5G and 6G) is becoming ever-increasing, which has beyond the capability of manual/semi-automatic management. 

Machine learning, especially deep learning, has shown its significant advantages on automating tasks in a wide variety of application areas, including pattern recognition techniques and game playing programs. The past few years have witnessed a deluge of research on applying deep learning techniques to achieve network automation in the telecommunication sector, including resource scheduling, route planning, anomaly detection, etc \cite{9060910}. This offers the opportunity to allow the increase in network complexity without the corresponding rise in network operational expenditures. Despite of the powerful capability of deep learning models on network automation in certain scenarios, network operators have little understanding on how these models make decisions or show certain behaviours \cite{DBLP:journals/corr/abs-2109-07239}. This is mainly due to the inscrutability nature of black-box deep learning models which are intrinsically hard to explain \cite{Rudin2019}. Because of the lack of transparency and trust of deep learning models, telecommunication operators are reluctant of having wide deployment in their networks but rather making them handle ``small'' tasks that they are able to control in case of unexpected model behaviours. This is contradictory to the expectation of pervasive artificial intelligence (AI) in 6G, and therefore, hindering the fast development of 6G solutions.

Network operators expect sufficient explainability and transparency of machine learning models, which can provide clear rationale for the decision-making process and surface any weakness of the process before unexpected behaviours may actually happen. With the ultimate goal of letting AI to autonomously manage the network, three key principles of creating safer AI \cite{10.5555/1671238} need to be considered in the process of developing machine learning models for network automation towards 6G. 
Below shows the details of the three principles:
\begin{itemize}
    \item The only goal of the robot is to maximize human values.
    \item The robot is initially uncertain about what those values are.
    \item Human behaviour provides information about human values.
\end{itemize}

To incorporate these principles in the design of machine learning models/agents for network automation, there needs to be a way that enables the agent to learn, from scratch, how human operators manage the network on a variety of network operation tasks, e.g., network configuration. In the process of handling a new network operation task, the agent should be able to leverage the knowledge learnt from how human operators handle similar tasks and perform the task in a similar way to realize human values. Human operators need to understand how the agent itself handles the network task and regulate the agent's behaviours by introducing intents and constraints.

\begin{table*}[]
\caption{The main contributions of existing work and the difference with this work}
\renewcommand\arraystretch{1.5}
\begin{tabular}{|C{1.5cm}|C{2cm}|C{2cm}|C{2cm}|C{2cm}|C{1.6cm}|C{1.5cm}|C{1.4cm}|}
\hline
 & \multicolumn{4}{C{10cm}|}{Network automation solutions}  & Transferability & Explainability & End-to-end solution \\ 
 \hline
\multirow{4}{*}{Existing work} & \multicolumn{2}{C{5.5cm}|}{Traditional methods – optimization, game theory, etc.} & \multicolumn{2}{C{4cm}|}{Model explainers}  & \xmark  & \cmark  & \cmark        \\ 
\cline{2-8}

                                & \multicolumn{1}{C{3cm}|}{Traditional methods – optimization, game theory, etc.} & \multicolumn{2}{C{3cm}|}{Deep learning models to solve traditional methods} & \multicolumn{1}{C{3cm}|}{Deep learning model explainers} & \xmark               & \cmark              & \xmark                   \\ 
\cline{2-8} 
                                & \multicolumn{2}{C{5.5cm}|}{Deep learning / transfer learning models}              & \multicolumn{2}{C{4cm}|}{Deep learning model explainers}                                                                             & \cmark               & \cmark              & \xmark                   \\ 
\hline
Our work                 & \multicolumn{4}{C{10cm}|}{A knowledge-powered human-understandable explainable AI framework}                                                                                                                                                             & \cmark               & \cmark              & \cmark                   \\ \hline
\end{tabular}
\label{tab:state-of-the-art}
\end{table*}

To achieve this purpose, in this article we propose a new knowledge-powered framework for network automation that can 1) effectively adapt to the changing and ever-increasingly complex environment of communication systems and 2) provide human-understandable explanation, in a step-by-step manner, of how the decision of a network automation task is reached.

\begin{figure*}
    \centering
    \includegraphics[scale=0.6]{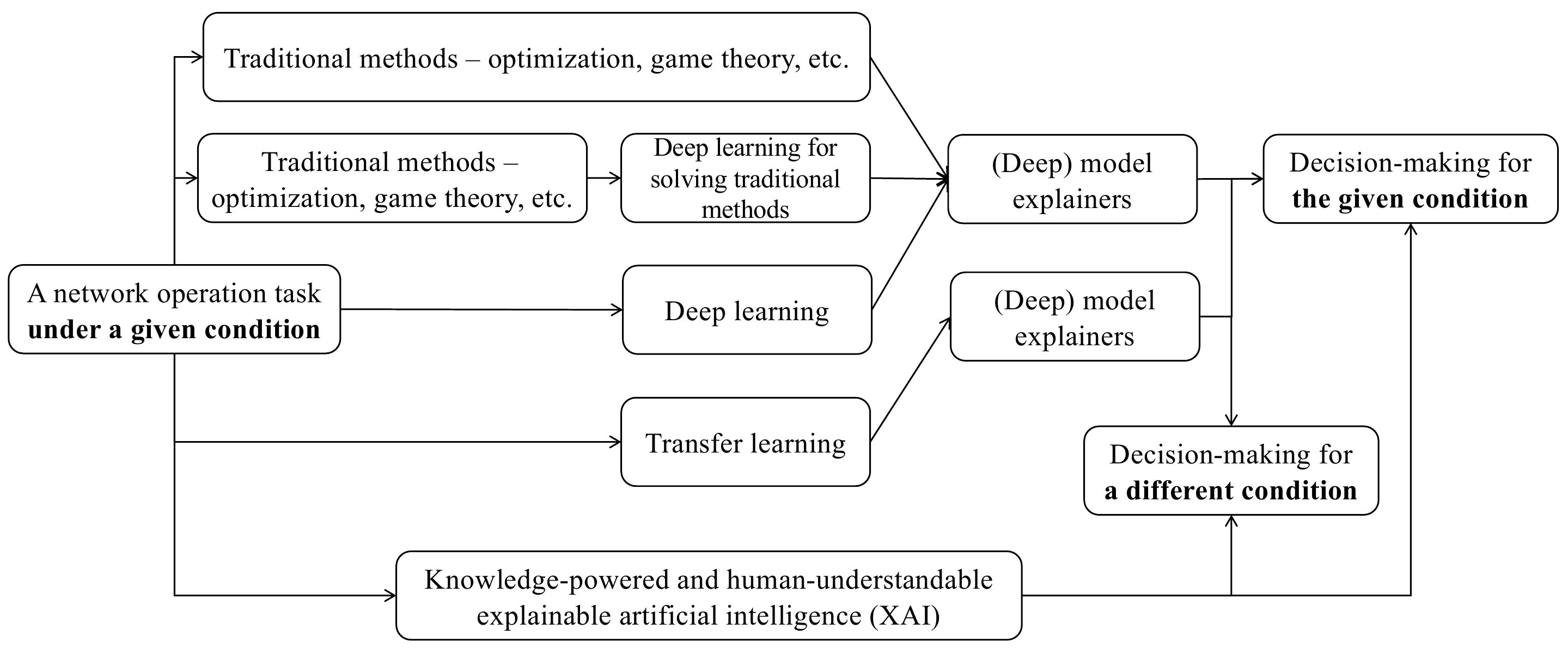}
    \caption{The workflow of existing work and our work in reaching an explainable decision making for network automation.}
    \label{fig:state-of-the-art}
\end{figure*}

\section{State-of-the-Art}
Traditional approaches were mainly to explicitly formulate network automation tasks as an optimization problem and then exploit mathematical programming to solve it to a certain level of optimality. Many network automation tasks such as routing and path selection problems are essentially a combinatorial optimization and integer programming problem. For such problems, exact algorithms based on exhaustive search are only useful for small-sized models, so approximation and heuristic algorithms were usually resorted to solve optimization problems. This set of methods was successful for relatively static and small-sized networks, but they do not very validate to modern complex wireless communication systems due to the ever-increasing size, high dynamics and unknown circumstances caused by emerging applications such as autonomous vehicles and flying drones. Many stochastic combinatorial optimization models \cite{9107260} were then built to tackle network management tasks in this kind of dynamic and uncertain environment. However, the success of related optimization solutions in the stochastic systems often requires solving a base optimization problem repeatedly until convergence for each optimization task. For each iteration, the base program exhibits the same model structure, but only differs in their input data. Deep learning techniques can be applied to many optimization problems by automatically detecting their heuristics based on training data, requiring less manual engineering than solvers optimized for a single problem or just one specific instance of a problem. Recently, there has been studies on using deep learning architectures based on graph neural networks to learn optimization solutions for combinatorial routing problems \cite{NEURIPS2018_9fb4651c}. But the research along this line is still in its infancy.

Apart from applying deep learning to solve optimization problems for network automation, many deep learning models were directly developed and used to solve network automation tasks \cite{8382166}. 
The main line of research is reinforcement learning. Its main idea is to find out which action will maximize the expected reward in the future, based on the current state of the environment. It has achieved remarkable results in the field of games and robot control. The technology has been transplanted to the network field to solve network automation problems, such as virtual network embedding \cite{9060910}, network functions virtualisation and placement \cite{8945291}, and resource management in network slicing \cite{WANG2019106}. However, the lack of modeling of the environment in model-free reinforcement learning methods makes the agent difficult to adapt the previously learned strategies to a new environment \cite{DBLP:conf/icml/KanskySMELLDSPG17}. As critical elements in wireless communication systems change, it becomes a completely new environment for the agent to handle. The model-free reinforcement learning agent must re-collect the interaction data with the environment and retrain the model to adapt to the new environment.

To sum up, there are two main issues of the state-of-the-art. One is the lack of efficient adaption to make the developed model applicable across changing environment of wireless communication systems. The other is that deep learning solutions are lack of transparency due to the black-box nature. Although many promising deep learning models have been developed, network operators are still reluctant to adopt them in real-world networking environment due to the above issues. Transfer learning has been leveraged to address the first issue, and for the second issue, studies on explainability of black-box deep learning models have been initiated.

The end-to-end solution proposed in this article has explicitly addressed the above two issues from a new angle, where an agent is developed to learn knowledge from network data collected against each network management practice. The knowledge learnt from various network settings can be accumulated and can be readily applied to complete a network task in a new network setting. The knowledge used to complete network tasks can also be used to explain how the task is accomplished. Table \ref{tab:state-of-the-art} summarizes the main contributions of existing work and the difference with our work in this article. Fig. \ref{fig:state-of-the-art} shows the workflow of existing work and our work in reaching an explainable decision making for network automation.

\section{An End-to-End Knowledge-powered Human-understandable Explainable AI (XAI) Agent for Network Automation}
\begin{figure*}[ht]
\center
\includegraphics[scale=0.55]{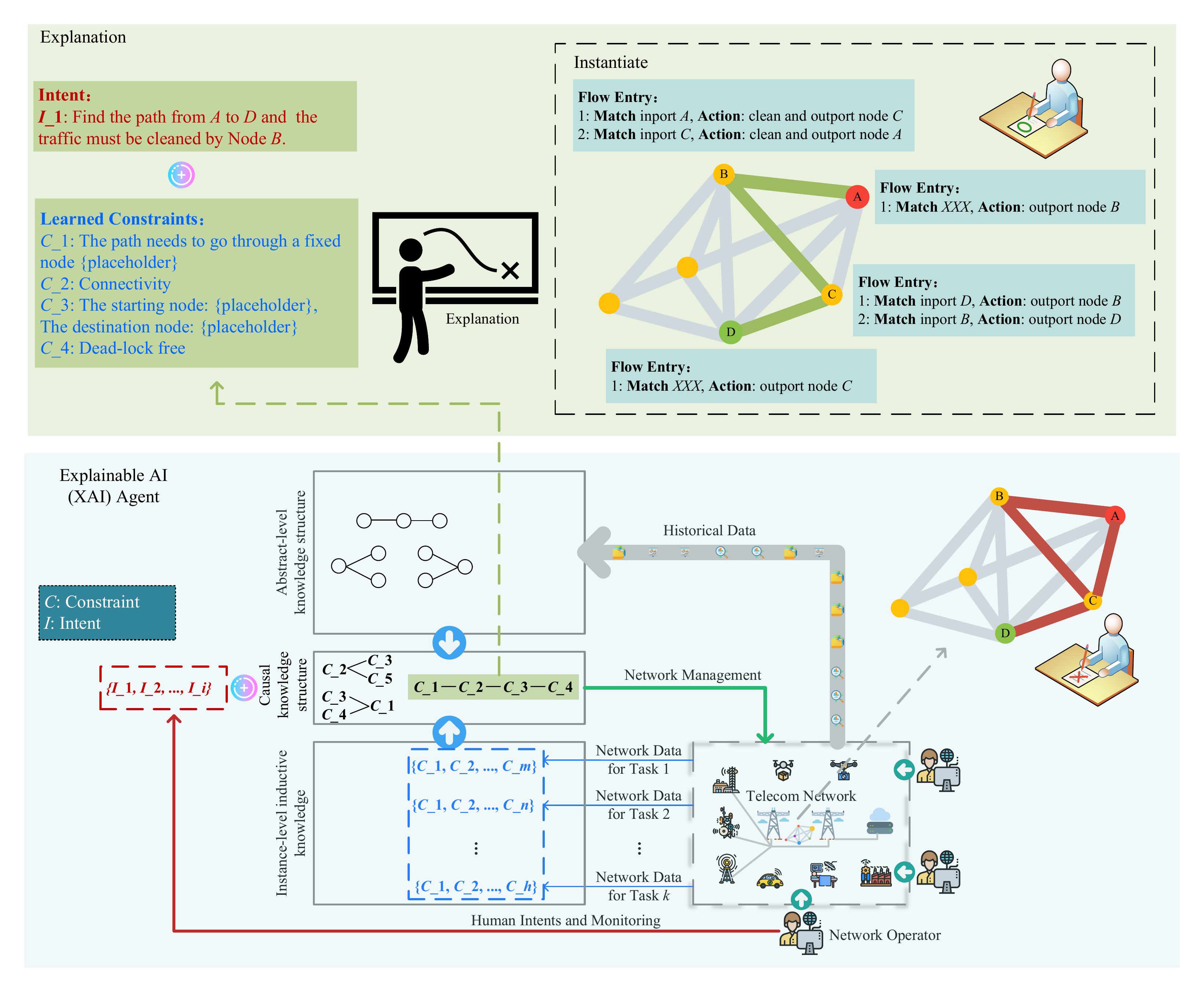}
\caption{The architecture of the proposed knowledge-powered human-understandable explainable artificial intelligence (XAI) agent for network automation.}
\label{fig:architecture_xai}
\end{figure*}

\begin{figure}[ht]
\center
\includegraphics[scale=0.095]{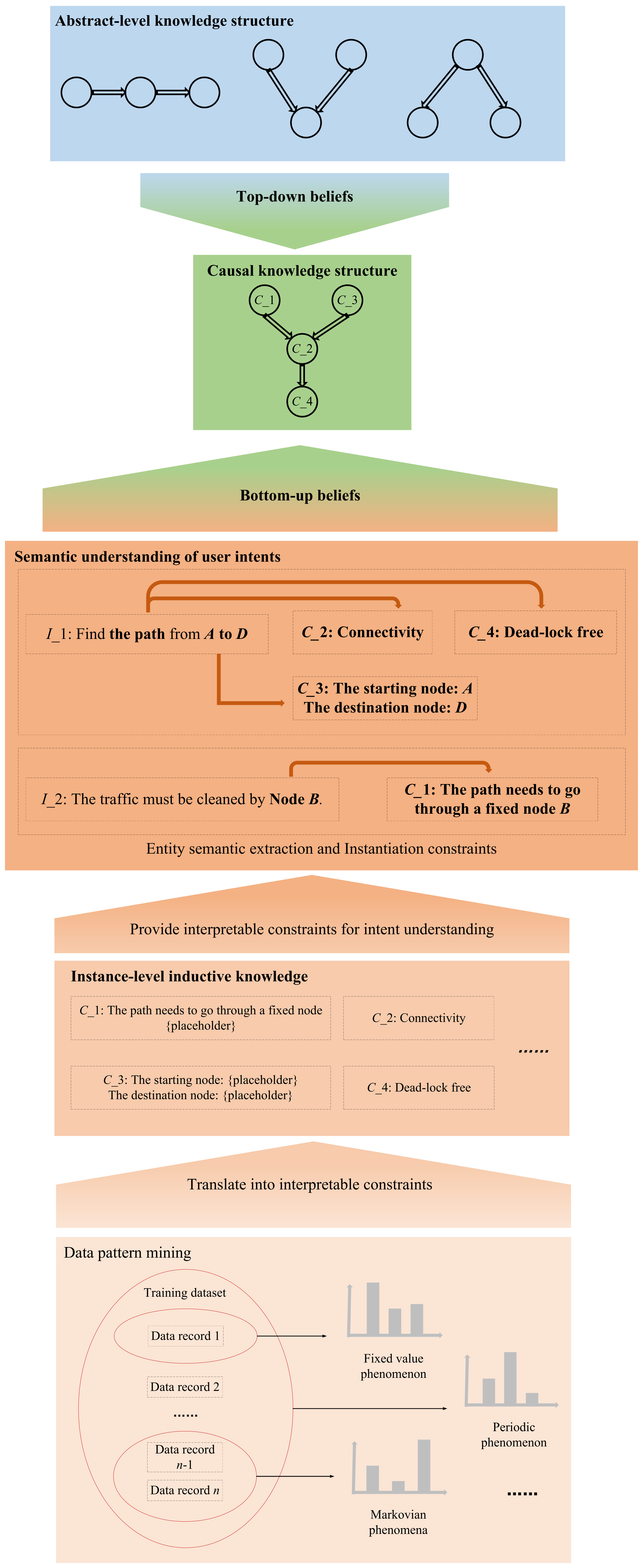}
\caption{The proof-of-concept model to support the proposed XAI agent.}
\label{fig:proof_of_concept_xai}
\end{figure}

The proposed explainable artificial intelligence (XAI), as shown in Fig. \ref{fig:architecture_xai}, is based on theory-based causal induction and Bayesian inference \cite{9472841}. It consists of four parts: 1) top-down abstract-level knowledge structure learning, 2) bottom-up instance-level inductive knowledge learning, 3) causal knowledge structure learning, and 4) human intents and monitoring. In what follows, each part will be elaborated, and how the proposed XAI agent can explain the decision making for network automation, in a human-understandable way, will be presented. Finally, the discussion of a proof-of-concept model to support the XAI agent, as shown in Fig. \ref{fig:proof_of_concept_xai}, will be carried out.

\subsection{Abstract-level knowledge structure learning} \label{sec:Abstract-level knowledge structure learning}
To enable learned knowledge from network operation practices applicable across different networks and/or network environment settings, we devise an \textit{environment-invariant abstract structure} that can be used to encode generalized knowledge of a network operation practice. Such practices include resource allocation and scheduling, traffic engineering, path planning, anomaly detection, just to name a few.
The environment-invariant abstract structure is essentially to express the various arrangements of constraints that match the definition of a network operation practice. For example, a linear structure can express the number of constraints (e.g., load balancing, connectivity, etc.) provided by a path planning practice and their linear arrangements.
The purpose of arrangement of these constraints is to provide the top-down theory for causal induction, which is the top-down belief about the task structure of a network operation practice.

\subsection{Instance-level inductive knowledge learning} \label{sec:Instance-level inductive knowledge learning}
Whilst learning an abstract structural understanding of a network operation practice, the agent shall interact with the communication system to learn which instance-level constraints significantly affect the performance of the network operation practice, as shown by the symbol $C$ in Fig. \ref{fig:architecture_xai}. In other words, we hope that causal events occurred in the network operation practice can be encoded with a constraint. A basic assumption is, yet general associative learning theory stands, i.e., causal relations will induce the state changes of the system, and non-causal relations do not. A likelihood term is learned for each constraint using a distribution. Note, this distribution can be a given distribution such as polynomial or a distribution fitted by a neural network. We strive to assess the likelihood of specific causal information that drives network practices. This associative likelihood is how likely the certain constraints will be associated with a causal event, given how often the constraints have been present in causal events in the past, without considering any abstract-level knowledge structure about the network practice. This instance-level learning provides bottom-up belief that reveals a task-invariant knowledge of which causal chains are more likely to cause a causal effect.

\subsection{Causal knowledge structure learning}
The abstract-level knowledge structure learning in Section \ref{sec:Abstract-level knowledge structure learning} provides environment-invariant abstract structures of a network operation practice. The instance-level inductive knowledge learning in Section \ref{sec:Instance-level inductive knowledge learning} provides the causal plausibility of the network operation practice. The goal of the causal knowledge structure learning is to pick up the action for network operation practices it believes has the highest chance of 1) being part of the desired solution for the network operation practice and 2) being causally plausible in the communication system environment. The processes of abstract-level knowledge structure learning and instance-level inductive knowledge learning, as well as the intents from network operators as shown by the symbol $I$ in Fig. \ref{fig:architecture_xai}, are integrated into a final 
causal knowledge structure posterior. This posterior is able to reflect the probability of the selected action in achieving the above goal. 
Once an action is selected to move towards the goal, the agent updates its beliefs of top-down abstract-level knowledge structure learning and bottom-up instance-level inductive knowledge learning. This iterative process makes the best decisions on network operation practices based on the current understanding of the environment, and then, it updates beliefs based on the observation results of instance-level inductive knowledge learning.

\subsection{The human-understandable explainability of decision-making}
Different from black-box deep learning models for making decisions, the proposed XAI agent leverages theory-based causal induction \cite{Griffiths_2009} through combining the top-down beliefs provided by the abstract-level knowledge structure learning, the bottom-up beliefs enforced by the instance-level inductive knowledge learning, and the intents introduced by network operators, and uses Bayesian inference, to make multiple attempts to learn the correct causal knowledge structure of making a network operation practice. According to the causal knowledge structure that has been found, the agent can trace back to the source to find and update its abstract structure to enable it to be applicable for the network operation practice across different settings and conditions of communication systems. The explainability of the decisions made by the agent is due to the strong fact that the causal knowledge structure is learned step-by-step by instance-level inductive knowledge learning, i.e., which constraints a network operation practice needs to follow in order to reach the final decision. This step-by-step causal learning is able to show a good understanding to users in a human-understandable way, e.g., how a routing decision was made. This explainability can be exhibited by the causal chain posterior at each learning step, and this causal chain posterior can be readily transformed into visual representations of how a network management operation such as traffic route is found given a set of constraints and human intents. For example, if two routing constraints, i.e., connectivity and load balancing, are learned from the traffic data in the process of interacting with the environment, the agent shall be able to conduct which constraint should be used first and which one should be the second, and eventually this causal chain will be demonstrated through its posterior. The corresponding explanation can be worked out using this causal chain posterior, including e.g. why a link is selected to form a traffic route and why a link is not.

\begin{figure*}[ht]
	\center
	\includegraphics[scale=0.14]{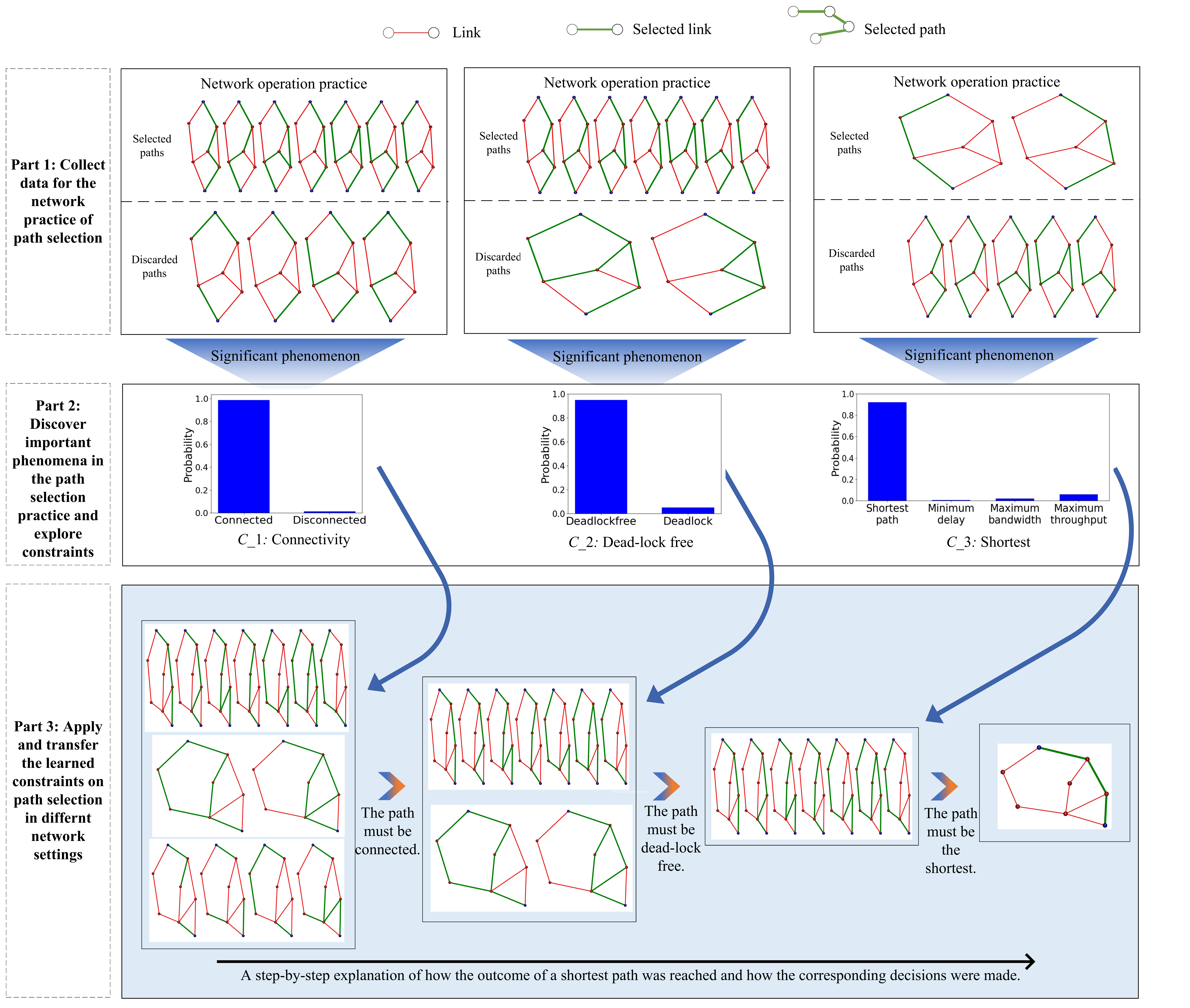}
	\caption{A case study of explaining the decision making for automated paths selection.}
	\label{fig:case_study}
\end{figure*}

\subsection{The discussion of a proof-of-concept model} \label{sec:The discussion of a proof-of-concept model}
The explainability of a network operation practice made by the proposed XAI agent is achieved with the help of causal knowledge structure that is obtained by merging top-down beliefs and bottom-up beliefs. As shown in Fig. \ref{fig:proof_of_concept_xai}, the proof-of-concept model discusses the general process of implementation. At the bottom part of the figure, the bottom-up belief first learns the salient phenomena of the network dataset, which imply the constraints behind the data. Essentially, the salient phenomena of data mainly show in three 
aspects: a single data point, adjacent data points, and the entire dataset. In terms of a single data point, the common phenomenon is a fixed value. The possible constraint behind the phenomenon may be that a certain node is constantly processing data. For example, the traffic passes through a virtual network function for cleaning. The label information of the virtual network function will be added in each traffic flow. For adjacent data points, it is necessary to infer whether there is a Markovian phenomenon in the dataset. The Markov phenomenon may imply that the sequence data has a pre- and post-constraint relationship. For example, in a path selection process, 
the selected adjacent links must be connected. The global information of the entire dataset may exhibit a periodic or non-periodic phenomenon such as dead-lock free paths. It means that the same link and/or node cannot be traversed repeatedly.

A variety of constraints can be mined by discovering significant phenomena in the training data. For the constraints learned from the data, the information about the environment in expression is replaced by a placeholder, and such learned constraints provide interpretation for the semantic understanding of user intentions. Entity recognition is performed through user intentions, and the semantics expressed by the entities are mapped to the constraints. Entities related to the environment are then assigned to the placeholder position of the constraint. For example, for the semantic understanding of user intents shown in the middle of Fig. \ref{fig:proof_of_concept_xai}, the user intention is to find a path from node $A$ to node $D$, and the traffic has to be cleaned at node $B$. The entities mentioned in the intent include the path, and the nodes $A$, $B$, and $D$. The path is mapped to the constraints of $C_2$ connectivity and $C_4$ dead-lock free. Since the nodes $A$, $B$ and $D$ are environmental entities, the mapped constraints need to be instantiated as $C_3$ (
The starting node: $A$, 
The destination node: $D$) and $C_1$ (The path needs to go through a fixed node $B$).

These four instantiated constraints constitute the bottom-up belief which contains all the constraint information used to complete the task described by the user intention. However, since there is an order of precedence among constraints, it requires top-down beliefs to supplement information about the environment-invariant abstract structure, which is the various arrangements of constraints that match the definition of a network task. The different arrangements of constraints are called causal knowledge structures.
In order to verify the accuracy of causal knowledge structures. We design two parameters $P$ and $R$. Before defining them, it is necessary to explain the concepts of target space and solution space. For a given task, the solution space contains all the possible solutions, and the solution which meets the user intention is an element in the target space. The target space belongs to the solution space. The essence of autonomous task execution is to use the causal knowledge structure to search in the solution space to get the target space. Each constraint in the causal knowledge structure can be regarded as a classification hyperplane that divides the solution space  into smaller subspaces. When the divided subspace is equal to the target space, the causal knowledge structure can correctly find the solution that satisfies the user intention. The parameter $P$ defines the proportion of the target space in a subspace, and the parameter $R$ indicates the percentage of the subspaces (derived from the constraints) in the target space. In an ideal situation, both $P$ and $R$ are equal to 1, meaning that the causal knowledge structure can correctly find a solution that satisfies the user intention. The parameters $P$ and $R$ measure the accuracy of the causal knowledge structure to perform network tasks. When deployed in a network, the causal knowledge structure can be interpreted as software defined networking (SDN) flow entries or virtual network functions.

The cross-environmental decision-making capabilities of the XAI agent are embodied into two aspects. One is that every constraint learned by the instance-level inductive knowledge is cross-environmental. In fact, a constraint can be treated more as a template, where the description of the environment is replaced by a placeholder. When a user proposes an intention, it is generally proposed for a network environment, such as ``find a path from node $A$ to $E$ in the current network''. To combine the intention with the cross-environment constraint, a constraint needs to be instantiated, that is, the placeholder is replaced with the entity related to the environment mentioned in the intent. Another embodiment is that the abstract-level knowledge structure provides structural information about constraints that are not related to the environment.  It is about the priority rules of constraints that need to be followed to complete a task.

\section{A Case Study}
Path selection is a typical application in 6G and network automation. For two arbitrary nodes in  Space–Air–Ground integrated networks, how to  select paths to optimize the performance in terms of efficiency, reliability and security of transmission is an important research problem \cite{8612450}. This section uses path selection as a typical example to intuitively demonstrate the feasibility of the proposed human-understandable XAI agent for network automation.

Fig. \ref{fig:case_study} shows a case study of our model. From top to bottom, it is divided into three parts, i.e., collecting data for the network practice of path selection, discovering important phenomena in the path selection practice data and explore constraints, and testing the usability of the learned causal knowledge structure under a different network topology setting. In the first part (Part 1) of data collection, each network operation practice corresponds to a set of selecting and discarding behaviors. The agent discovers characteristics of these data to infer the reasons behind the behavior and forms its understanding of the important phenomenon of the data. The agent learns instance-level constraints by comparing the difference between the selected and discarded paths. For example, in the first network operation practice in the upper left corner, the connected paths are selected, and the disconnected paths are discarded. Therefore, the connectivity, denoted as $C_1$, is an important constraint in this practice. Constraints are expressed using the probability distributions shown in the second part (Part 2) of the figure. Following the process of learning $C_1$, the other two constraints, $C_2$ dead-lock free and $C_3$ shortest path, can be readily obtained.

These constraints are the factors of network operation practices that occur in the paths selection process. They each independently explain why the path was selected in that way. It is worth noting that the constraints are decoupled from the environment setting. The reasons for events that occur during the execution of the task can be seen as the knowledge of completing that task. These instance-level knowledge form a causal knowledge structure that is essentially a way of expressing the priority among constraints. The determination of this priority comes from the top-down belief about the task structure of a network operation practice. In this case study, the three constraints are inferred to be a linear relationship from the high to low priority $C_1 \rightarrow C_2 \rightarrow C_3$.

The bottom part (Part 3) is the experiment about the applicability and transferability of the learned causal knowledge structure. Different from the network topology where the previous training data is collected, the agent uses the learned causal knowledge structure under the new network topology setting and manages to complete the task of path selection autonomously. Each constraint is actually a classification hyperplane, and those paths that meet the constraints will be retained, otherwise they will be eliminated. According to the linear causal knowledge structure $C_1 \rightarrow C_2 \rightarrow C_3$, the agent first uses the $C_1$ ``connectivity'' constraint to eliminate invalid paths that are not connected. Then it leverages the $C_2$ constraint to keep the paths without loops, and finally it uses the $C_3$ ``shortest'' constraint to select the shortest path. 

\begin{figure}[ht]
	\center
	\includegraphics[scale=0.45]{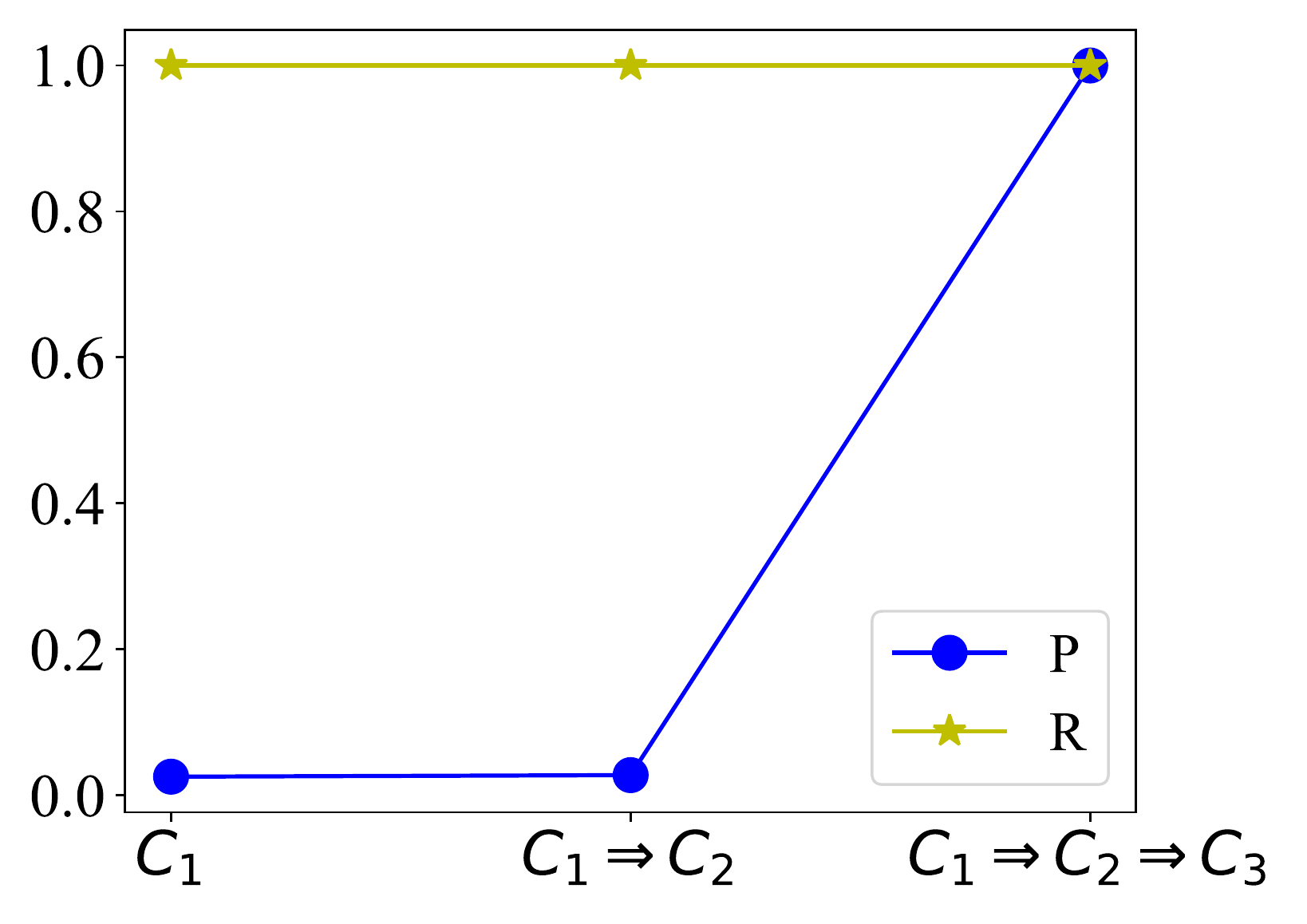}
	\caption{The parameters $P$ and $R$ indicate the availability of the causal knowledge structure.}
	\label{fig:PandR}
\end{figure}

Fig. \ref{fig:PandR} is about the evaluation of the parameters $P$ and $R$ mentioned in Section \ref{sec:The discussion of a proof-of-concept model}, which shows that the agent uses the causal knowledge structure to perform path selection in a step by step manner. The abscissa represents the three steps that the agent performs the task of finding the shortest path. The first step uses
$C_1$ to find all the connected paths in the network. The next two steps are to superimpose constraint $C_2$ and constraint $C_3$ on the causal knowledge structure respectively on the basis of the previous step. In each step, the measurement $R$ is equal to 1, indicating that at each step, the constraint does not remove the target path by mistake.  Moreover, the measurement $P$
gradually approaches to 1, indicating that the paths selected  are  more in line with the requirements of the shortest paths. In the last step, both the $R$ and the $P$ are equal to 1 , which demonstrates that the agent finds the shortest path accurately.

Causal knowledge structure transparently provides a clear rationale for the decision-making process since the instance-level constraints are clearly stated. It
provides a human-understandable explanation, in a step-by-step manner, of how the outcome of a network automation task is reached and how the corresponding decisions are made. The purpose of this case study is to visually illustrate the mechanism of inferring constraints from the network data and its role in the process of task execution.

\section{Research Challenges and Open Issues}
Although many studies have been reported on network automation and explainability, there are still many research challenges. In this section, we discuss some challenges and open issues which can guide further research on this topic.

\begin{itemize}
    \item \textit{Ethics-responsiveness}. Network automation, on the one hand, is to let the network itself make decisions / perform actions automatically based on the knowledge they learned from the network data. On the other hand, we need to ensure that the decisions do not make any harmful impact on the network services. For example, a decision of allocating computing resources to certain applications in a bias way would be treated as a harmful impact \cite{leslie2019understanding}. This is particular important in the so-called Zero-Touch Networks \cite{8994961} where all operational processes and tasks are expected to be executed autonomously. In this case, the decision-making that enables network automation has to be responsible and trustworthy. Embedding ethics-responsiveness into the design of AI models is a challenge. It is a multi-disciplinary research that requires expertise of both Computer Science and Social Science. For example, it is not straightforward to formalize an ethical and moral theory such as divine command theory and embed it into an AI model so that the model can make decisions to take morally right actions on network tasks.
    \item \textit{Computationally efficient explainability}. Explainability is one of the many ways to ensure the responsiveness and trustworthiness of a decision-making. It can allow the monitoring system to understand why a decision was made, hence assessing whether there is any potential ethical issues, e.g., bias. Explainability can be realized through various ways. Some studies tried to unveil the internal process of a learning model. Some work managed to explain the relationship between input features and output decisions, i.e., which features have more weights towards the decision-making. Some studies explained how a decision was made, step-by-step, by searching a space. Existing explanation methods are computationally inefficient and require significant hyper-parameter tuning. It is non-trivial to shorten the convergence time in different ways of achieving explainability.
    \item \textit{Automated network assessment and monitoring}. For network automation, on the one hand, operation decisions are automatically made. On the other hand, there has to be an automated assessment that can monitor whether the decision was made correctly. If it is deviated, appropriate actions need to be taken automatically. For example, if a bias decision was made, more diverse data may need to be introduced to re-train the model, or a human operator may need to get involved to ``teach'' the model in order for it to behave ethically. Enabling automated assessment of AI models to cover all the potential test cases including the corner cases is a hard problem from the perspective of software testing. Guaranteeing free of ethical issues of automated assessment is also worth of investigation.
    \item \textit{Closed-loop control}. Network automation requires a closed-loop control from the data collection, data analysis, to the decision making. Currently, the three phases are not streamlined, instead they are separated. For example, data are still collected separately and at the time of collection, it often does not consider the requirement of data analysis and decision making. This may result in low performance of the data analysis and decision making. More importantly, the control loop is not closed, hence cannot make the aim of network automation. Closing the control loop is challenging, which requires comprehensive coordination and optimization amongst data collection, data analysis and decision making phases. It is also non-trivial to have appropriate channels for feedback between each of the above three phases. In addition, it is crucial for AI models in each phase to consume the feedback and automatically update themselves by virtue of e.g. automated machine learning (AutoML).
\end{itemize}

\section{Conclusion}
This article proposed a knowledge-powered architecture for network automation with human-understandable explainability. The causal knowledge that can be used to automatically complete a network task, were generated by 1) the instance-level knowledge learnt from the network data for a given task/condition, 2) the abstract-level knowledge learnt from historical network data, and 3) the intents from network operators. The proposed knowledge-powered architecture is able to provide a human-understandable and step-by-step explanation of how the outcome of a network automation task was reached and how the corresponding decisions were made. In addition, the learnt knowledge can be applicable across different networks and/or network settings. A case study was developed to demonstrate the feasibility of the proposed knowledge-powered architecture for automatically performing path selection tasks. At last, we provided a list of research challenges and open issues that can be useful to the community for carrying out further research.

\bibliographystyle{IEEEtran}
\bibliography{Ref_bib}

\begin{IEEEbiographynophoto}{Yulei Wu}
[Senior Member] is a Senior Lecturer with the Department of Computer Science, College of Engineering, Mathematics and Physical Sciences, University of Exeter, UK. He received the Ph.D. degree in Computing and Mathematics from the University of Bradford in 2010. His research interests include networking, IoT, edge intelligence, privacy and trust, and AI and ethics.
\end{IEEEbiographynophoto}

\begin{IEEEbiographynophoto}{Guozhi Lin}
is now a Ph.D. candidate of the University of Chinese Academy of Sciences. He received the B.Sc. degree in computer science and technology from South China Normal University in 2016. His research interests are mainly in intent-based networks, network automation and reinforcement learning.
\end{IEEEbiographynophoto}

\begin{IEEEbiographynophoto}{Jingguo Ge}
is currently a Professor with the Institute of Information Engineering, Chinese Academy of Sciences, and a Professor with the School of Cyber Security, University of Chinese Academy of Sciences. He received the Ph.D. degree in computer system architecture. His research focuses on computer network architecture and cyber security.
\end{IEEEbiographynophoto}


%





\ifCLASSOPTIONcaptionsoff
  \newpage
\fi

\end{document}